%% file: E6July18.tex
\font\eightrm=cmr8 at 8pt
\documentclass{oxarticle}
\usepackage{amsmath, amssymb}
\usepackage{graphics, setspace}
\usepackage{epstopdf}
\usepackage{bm}
\usepackage{tikz}
\usepackage{xcolor}

\input{Johnmacros29}

\newcounter{orange} 
\newcounter{apple} 
\addtocounter{apple}{1}
\newcounter{grape} 
\addtocounter{grape}{1}
\newcommand{\numberhere}{TheEPapersJuly12\;}
\newcommand{\articlenumber}{E6July18}

\usepackage{color}

\newcommand{\mathsym}[1]{{}}
\newcommand{\unicode}[1]{{}}

\usepackage{tikz}
\newcommand*\circled[1]{\tikz[baseline=(char.base)]{
            \node[shape=circle,draw,inner sep=2pt] (char) {#1};}}
  
\begin{document}

 \begin{center}
{ \LARGE
The Supersymmetric Standard Model, combined with a special \EI, yields a new kind of SUSY mass splitting (E6) \\
[1cm]}  
%

\renewcommand{\thefootnote}{\fnsymbol{footnote}}

{\Large John A. Dixon\footnote{jadixg@gmail.com, john.dixon@ucalgary.ca}\\Physics Dept\\University of Calgary 
\\[1cm]}  

 \end{center}
 \Large   

 \begin{center} Abstract
 \end{center}
 \EI s, with Lorentz invariance, have been found in the BRS cohomology of SUSY in 3+1 dimensions.  Until recently,  it was generally accepted that no such objects could exist.  It is shown here that the \SSM\  (``the SSM'')  can be coupled to a special  \EI\ to form a new model, which we call the \XM\ (``the XM'') .  This \XM\ continues to generate the usual \vev\    (``VEV'') that breaks gauge symmetry from \sg\ to \bsg.  But now, from the same VEV, the \XM\ also generates SUSY violating mass splitting of the neutral ``ZX'' sector at  tree level.  This is possible because the \EI\  changes the algebra of SUSY.  This ``ZX'' sector is flavour neutral, which might explain the suppression of flavour changing neutral currents observed in experiments.  The XM is governed by a very restrictive Master Equation, so there should be a very small number of parameters in it.

  \section{Introduction}
  \subsection{The motivation for this paper}
  
In some recent papers, it has been  shown that the BRS cohomology of supersymmetry (``SUSY'') is non-trivial \ci{E1,E2,E3,E4,E5,E6,E7,E??}.  In fact it admits Lorentz invariant objects, which we call ``\EI s''.  These \EI s necessarily contain pseudofields, as well as ordinary fields.  The pseudofields are the sources for the BRS variations of the component fields.   That means that when we take a SUSY theory and add an  \EI, we generate a theory that is no longer really a SUSY theory, because the pseudofields change the BRS algebra of SUSY in important ways.  The \EI s also necessarily contain an unusual representation of SUSY.  This is the \cdss\ multiplet (``the CDSS multiplet''), which contains two spinors and a vector field\footnote{The CDSS multiplet is just a chiral multiplet with an extra dotted index $\dot \a$. That index $\dot \a$ behaves inertly under SUSY transformations.  But the  presence of $\dot \a$  changes the character of the component fields of the chiral multiplet in important ways: the scalar becomes a spinor, the spinor becomes a vector, and the auxiliary becomes another spinor. It will be treated more fully in \ci{E7}, where it will be crucial for the mass splitting in the ZX  sector.  Its action is displayed below in section \ref{CDSSsection}.  Including  it in the theory  is the simplest way to form a Lorentz invariant \EI, given the results in \ci{E2}.}.  

SUSY theories have always been marred by the difficulty of finding a suitable way to break the symmetry that gives rise to degenerate masses within SUSY multiplets.  Spontaneous breaking of SUSY is possible, but it has numerous problems which have been extensively examined over the last fifty years.  Explicit soft SUSY breaking has also been carefully examined for many years, The  motivation for it is obvious, but it  has no satisfactory theoretical origin.  Anyways, it has too many parameters to be useful  \ci{ D’Onofrio}. The consensus seems to be that  the known methods to generate mass splitting for the  SUSY supermultiplets are awkward, and unsatisfactory for comparison with experiment   
\ci{Allanach:2024suz,D’Onofrio,haberetal,xerxes,weinberg}.  Moreover, so far, no evidence of SUSY has been found    \ci{ D’Onofrio}. 

The hope of this paper is that the XM might provide a sufficiently clear, well defined, and unique  set of predictions, so that SUSY could be useful as an explanation of the nature of the particle spectrum.

So, in  this paper, we look at the \XM\ (``the XM'')  for an alternative way to split the degenerate masses within SUSY multiplets.   The XM has a small number of parameters, and it
appears to give rise to mass splitting without any need for 
 spontaneous breaking of SUSY, or any need for explicit `soft breaking' of SUSY. 

As we shall see below, the mass splitting in the XM originates in the neutral ``ZX  sector''.  This ``ZX  sector'' consists of  the Z vector boson and its spinor and scalar superpartners, and  the CDSS vector boson X and its spinor superpartners.   	The ZX  sector is flavour diagonal, so mass mixing inside it does not differentiate between the masses of the quarks and lepton generations.  Also the ZX sector is the only sector where mass mixing splits SUSY at tree level.  So these two features together  might explain why flavour changing neutral currents  are experimentally observed to be suppressed. 

 What is clear is that at tree level, the quark and lepton multiplets, and the Higgs multiplet, and the photon multiplet,  and the $W^+,W^-$  vector multiplets, do not get any change to their supersymmetric mass matrices.  They all retain supersymmetric mass degeneracy at tree level.  The only violation of supersymmetric mass degeneracy, at tree level, occurs in the ZX  sector. This can all be easily seen by looking at the various parts of the action below in this paper. 
 
 But it is also clear that the tree level supersymmetry mass degeneracy violation, in the ZX  sector, will have an effect at one loop on all the other particle mass matrices.  What is not clear is whether the model really makes sense, and for that it is essential to look carefully at the ZX  sector.

The ZX  mixing is rather complicated, and so the next step, after this paper, is to calculate the prediction for the mass splitting of the ZX  sector in the XM, as a function of the VEV and of the various coupling constants.  These will include the new dimensionless coupling constant $g_x$ that couples the \EI\ to the SSM, and the independent mass $m_x^2$ of the CDSS multiplet in section \ref{CDSSsection} below. If this mass splitting makes sense, that will be significant.  If it does not make sense, we can wonder, why not?  The hope of this paper is that these questions can be answered, starting with \ci{E7}.

\subsection{The \SSM\ and the \Sei:}
 In this section, we review the properties of our version of the SSM. It contains a singlet J in addition to the usual H and K doublet fields that are needed to give masses to both up and down quarks. We also include three right handed neutrinos in the present model, so that the  CKM matrices for the electrons and the neutrinos are like those of the quarks. It is necessary to be definite about what model we start with, because the details of the superpotential are crucial to the considerations in sections \ref{countingsection}, \ref{constraintsection}  and  \ref{generatorsection} below. These assumptions are summarized in Table \ref{quantumnumbertableformatter}.
 
 The first three lines summarize our notation for the superpartners of a given chiral multiplet, for the case of the lepton doublet $L^{ip}$.  The notation for the other chiral multiplets of the SSM is similar.

\be
\vspace{.2cm}
\begin{tabular}{|c|c|c
|c|c|c
|c|c|c|}
\hline
\multicolumn{8}
{|c|}{ \bf Table 1}
\\
\hline
\multicolumn{8}
{|c|}{ \bf SU(2)  Scalar,  Spinor and Auxiliary 
  $Lepton$ Doublet   Fields}
\\
\hline
\hline
{\rm Field} & Y 
& {\rm SU(3)} 
& {\rm SU(2)} 
& {\rm Flavours} 
& {\rm Baryon} 
& {\rm Lepton} 
& {\rm Dimension} 
\\
\hline
$ L^{pi} $& -1 
& 1 & 2 
& 3
& 0
& 1
& 1
\\
\hline
$ \y_{L\a}^{pi} $& -1 
& 1 & 2 
& 3
& 0
& 1
& $\fr{3}{2}$
\\
\hline
$ F_{L}^{pi} $& -1 
& 1 & 2 
& 3
& 0
& 1
& 2
\\
\hline
\multicolumn{8}
{|c|}{ \bf  Scalar Fields for more  SU(2)  Doublets}\\
\hline
$ Q^{cpi} $ & $\fr{1}{3}$ 
& 
3 &
2 &
3 &
 $\fr{1}{3}$
& 0
& 1
\\
\hline
$H^i$ 
& -1 
& 1
& 2
& 1
& 0
& 0
& 1
\\
\hline
$K^i$ 
& 1 
& 1
& 2
& 1
& 0
& 0
& 1
\\
\hline
\multicolumn{8}
{|c|}{ \bf Scalar Fields for the SU(2) Singlets  
 }
\\
\hline
$P^p$ & 2 
& 1
& 1
& 3
& 0
& -1
& 1
\\
\hline
$R^p$ & 0 
& 1
& 1
& 3
& 0
& -1
& 1
\\
\hline
$T_c^p$ & $-\fr{4}{3}$ 
& ${\ov 3}$ &
1 &
3 &
 $-\fr{1}{3}$
& 0
& 1
\\
\hline
$B_c^p$ & $\fr{2}{3}$ 
& ${\ov 3}$ 
&
1 &
3 &
 $- \fr{1}{3}$
& 0
& 1
\\
\hline
$J$
& 0 
& 1
& 1
& 1
& 0
& 0
& 1
\\
\hline
\end{tabular}
\la{quantumnumbertableformatter}
\ee

In Table \ref{gaugeandXnames} we summarize our notation for the gauge particles and for the X multiplet. 

\be
\vspace{.2cm}
\begin{tabular}{|c|c|c
|c|c|c
|c|c|c|}
\hline
\multicolumn{8}
{|c|}{ \bf Table 2}
\\
\hline
\multicolumn{8}
{|c|}{ \bf SSM, The U(1) Gauge Fields}
\\
\hline
{\rm Field} & Y 
& {\rm SU(3)} 
& {\rm SU(2)} 
& {\rm Flav} 
& {\rm Bar} 
& {\rm Lep} 
& {\rm Dim} 
\\
\hline
$V_{\dot \a \b}$ & 0
& 1
& 1
& 1
& 0
& 0
& $1$
\\
\hline
$\lam_{ \a  }$ & 0
& 1
& 1
& 1
& 0
& 0
& $\fr{3}{2}$
\\
\hline
$D$ & 0
& 1
& 1
& 1
& 0
& 0
& $2$
\\
\hline
\multicolumn{8}
{|c|}{ \bf SSM, The SU(2) Gauge Fields}
\\
\hline
$V^{a}_{ \a \dot \b}$ & 0
& 1
& 3
& 1
& 0
& 0
& $1$
\\
\hline
$\lam^{a}_{ \a  }$ & 0
& 1
& 3
& 1
& 0
& 0
& $\fr{3}{2}$
\\
\hline
$D^{a}$ & 0
& 1
& 3
& 1
& 0
& 0
& $2$
\\
\hline
\hline
\multicolumn{8}
{|c|}{ \bf SSM, The SU(3) Gauge Fields}
\\
\hline
$V^{A}_{ \a \dot \b}$ & 0
& 8
& 1
& 1
& 0
& 0
& $1$
\\
\hline
$\lam^{A}_{ \a  }$ & 0
& 8
& 1
& 1
& 0
& 0
& $\fr{3}{2}$
\\
\hline
$D^{A}$ & 0
& 8
& 1
& 1
& 0
& 0
& $2$
\\
\hline
\multicolumn{8}
{|c|}{ \bf XM, The X Multiplet Fields}
\\
\hline
{\rm Field} & Y 
& {\rm SU(3)} 
& {\rm SU(2)} 
& {\rm Flav} 
& {\rm Bar} 
& {\rm Lep} 
& {\rm Dim} 
\\
\hline
$\f_{\dot \a}$ & 0
& 1
& 1
& 1
& 0
& 0
& $\fr{1}{2}$
\\
\hline
$X_{ \a \dot \b}$ & 0
& 1
& 1
& 1
& 0
& 0
& $1$
\\
\hline
$\c_{\dot \a}$ & 0
& 1
& 1
& 1
& 0
& 0
& $\fr{3}{2}$
\\
\hline
\end{tabular}
\la{gaugeandXnames}\ee

\subsection{Here is the superpotential for our version of the \SSM:}

\la{SSMsuper}

The  form for the superpotential, after the shifts $H^i\ra H^i + m h^i $ and $K^i \ra K^i + m k^i $, to remove the VEVs,  is 
\be
\cP_{\rm SSM} = 
g_{H} \fr{1}{\sqrt{2}} \e_{ij}( H^i + m h^i  )( K^j + m k^j )J   - g_{H} \fr{1}{\sqrt{2}}  m^2 \textcolor{green}{\circled{J}} 
\eb+
p_{pq} \e_{ij} L^{p i} ( H^j + m h^j  ) P^q
+
r_{pq} \e_{ij} L^{p i} ( K^j + m k^j )R^q
\eb
+
t_{pq} \e_{ij} Q^{c p i} ( K^j + m k^j ) T_c^q
+
b_{pq} \e_{ij} Q^{c p i} ( H^j + m h^j  )  B_c^q
\la{vacuumpotmassive}
\ee 
The term  \textcolor{green}{\circled{J}} breaks SU(2) $\times$ U(1) to U(1), but leaves SUSY unbroken.  This form contains the VEVs of the fields $H,K$, and a shift so that the VEV of the fields in this form is now zero. The VEVs $m h^i $ and $m  k^j$  are determined so that the linear terms in fields are removed:
\be
  \e_{ij}   h^i   k^j    =    1 
\ee

The following set of equations (\ref{doublets}) summarize the   names of the left handed doublets after the \sbgs.
\be 
 H^{i}=
\begin{bmatrix}
H^0   \\      H^-  \\  
\end{bmatrix}
;\; K^{i}=
\begin{bmatrix}
K^+   \\     K^0    \\  
\end{bmatrix}
;\; L^{pi}=
\begin{bmatrix}
 N^p    \\      E^{p}    \\  
\end{bmatrix}
;\;Q^{cpi} =
\begin{bmatrix}
   U^{c p}    \\  D^{c p}       \\     
\end{bmatrix}
\la{doublets}
\ee

We can remove some clutter by using the following abbreviated form for the superpotential 
  \be
\cP  = \lt \{t Q K T +r L K R+b Q H B+p L H P
+g_{H} \fr{1}{\sqrt{2}} H K  J
-
g_{H} \fr{1}{\sqrt{2}} m^2 J 
\rt \}
\ee

\subsection{The Counting Type Solution of the Constraint}
\la{countingsection}

In practical terms, for the present example, the
  easiest way to solve the constraint equation for an \EI\ is to find a counting operator $N_{\rm EI}$ such that
  \be
 N_{\rm EI}  \cP[A]  =0
\la{countingform}
\ee
where
\be
 N_{\rm EI}= \sum_i n_i N_i[A] \equiv \sum_i n_i A^i \fr{\pa}{\pa A^i}  
\ee
We need to express this in terms of the chiral  fields of the SSM, of course.
In this paper we will not try to discuss at length the origin and reasons for the form of the \EI s. This, rather complicated, discussion\footnote{In its simplest form, for present purposes, the result of \ci{E2} is that an \EI\ can be made in a SUSY theory, from an expression of the form
$\cE_{\rm Special} = \Ct  \f^{\dot \a}  \oy_{i \dot \a} ( t^i_j  A^j  + t^i  m )\la{nextEIspecseq}
$, where the constraint equation has the form
$d_2 \cE_{\rm Special} =0
$.  Here the constraint for the \EI s is of the form:
$d_2 =   \oC_{\dot \a} \sum_i \fr{\pa   \cP}{\pa A^i} \fr{\pa}{\pa \oy_{i \dot \a }} 
$. So 
we want to have matrices $t^i_j,  t^i$ such that
$  \fr{\pa \cP}{\pa A^i} ( t^i_j  A^j  + t^i  m )=0
 $ where the superpotential of the theory is 
$ P = g_{ijk} A^i A^j A^k+ m g_{ij} A^i A^j  + m^2 g_{i} A^i  
$. This is equivalent for simple cases to finding a solution to the equation (\ref{countingform}) above.  The extension of the spectral sequence argument in \ci{E2} to the inclusion of masses, and the CDSS, and the gauge theory, will be included in \ci{E??}.  No new issues are expected there, and the arguments in \ci{E2} should not be difficult to extend to include those features.}
 can be found in \ci{E1,E2,E3,E4,E5,E6,E7,E??}.
The result of all that discussion is summarized in the action terms in section  \ref{newtermsforXM} below.

\subsection{Satisfying the Constraint for the SSM, as set out above}
\la{constraintsection}

 The version of the SSM that we use here includes two Higgs supersymmetric chiral doublets $H^i$ and $K^i$,  as is usual in order to be able to give masses to both the up and down quarks (and the electrons and neutrinos too).  It also includes a neutral singlet supersymmetric chiral Higgs multiplet J which makes the superpotential a natural generator for gauge symmetry breaking from  \sg\ to \bsg.

If we choose   (\ref{countingform}) as a way of finding a solution to the constraint equation for an \EI, then we want to ensure that the solution survives the shift of scalar fields that results from a VEV and gauge symmetry breaking. More important than that, is that the theory should exhibit some splitting of the mass spectrum of SUSY {\bf at tree level}.  When we look at the form of the \EI s (they are written out explicitly in section \ref{newtermsforXM} below), it becomes obvious that this means that we want the \EI\ to include terms that arise from the Higgs fields H and K, because these get the shifts that create mass terms:
\be
H^i \ra H^i  + m h^i
\ee
\be
K^i \ra K^i  + m k^i
\ee

 Then the question is how to do that.  

Consider the form of the superpotential for the model here, which is, in abbreviated form: 
  \be
\cP  = \lt \{t Q K T +r L K R+b Q H B+p L H P
+g_{H} \fr{1}{\sqrt{2}} H K  J
-
g_{H} \fr{1}{\sqrt{2}} m^2 J 
\rt \}
\ee  
We divide this into several parts
  \be
\cP =\cP_{\rm Matter}  +  \cP_{\rm HJK}   + \cP_{\rm m^2 J} =\cP_{\rm Matter}  +   \cP_{\rm Higgs} 
\ee
where  \be
\cP_{\rm Matter}  = \lt \{t Q K T +r L K R+b Q H B+p L H P
\rt \}
\ee   \be
  \cP_{\rm HJK}    = \lt \{ +g_{H} \fr{1}{\sqrt{2}} H K  J
\rt \}
\ee   \be
   \cP_{\rm m^2 J}  = \lt \{ -
g_{H} \fr{1}{\sqrt{2}} m^2 J 
\rt \}
\ee  
We can combine the last two into one expression:
\be
\cP_{\rm Higgs} =  \cP_{\rm HJK}   + \cP_{\rm m^2 J} 
 \ee
 
 Now we note the following simple counting relations:

 \be
N_{\rm Special} \cP =0
  \ee
\be
N_{\rm Leptons}\cP =0
\ee
\be
N_{\rm Quarks}\cP =0
\ee
where we define
  \be
N_{\rm Special} =
  N_H-
 N_K
 +N_T
- 
  N_B
- N_P
+ 
N_R
\la{nspecial1}
 \ee
   \be
N_{\rm Leptons}=  N_L - N_R - N_P 
\la{Nlepton}
 \ee
 \be
N_{\rm Quarks}=  N_Q - N_T - N_B 
\la{Nquark}
 \ee
 
 It is true that
  \be
N =  (N_H - N_K  )\cP_{\rm Higgs} =0
 \ee 
 but there are problems with the equations:

 \be
\lt \{n_1 (N_H+
N_K)  + n_2 N_J\rt \}\cP_{\rm Higgs}=0
\ee
because these imply that
\be
 \lt \{ 2 n_1   + n_2  \rt \}  \lt \{ +g_{H} \fr{1}{\sqrt{2}} H K  J
\rt \}=0
\ee  
and
 \be
\lt \{  n_2  \rt \}     \lt \{ -
g_{H} \fr{1}{\sqrt{2}} m^2 J 
\rt \}=0
 \ee
Assuming that $g_H \neq 0$,
it follows that either $m^2 =0$ or $n_1   = n_2=0$.  That means that, assuming we want our \EI\ to exist for the spontaneously broken theory $m\neq 0$, we cannot get an \EI\  that satisfies the constraint equation unless we exclude both $(N_H+
N_K)$ and $ N_J$ from our choice of $ N_{\rm EI}$ to solve equation (\ref{countingform}).   A related difficulty was experienced in the paper \ci{E1}.

What this means is that we can find solutions of the form
  \be
N_{\rm EI} =
n_s N_{\rm Special} +
n_l N_{\rm Leptons}+
n_q N_{\rm Quarks} 
 \ee
 any of which, for a nonzero value   $n_s$, will generate masses that split SUSY, at tree level, in the neutral ZX  sector.  But we cannot generate an \EI\ if we choose to keep $n_1 (N_H+
N_K)  + n_2 N_J$ as part of our $N_{\rm EI} $, because those counting operators do not generate solutions to the constraint when the theory breaks gauge symmetry. 

For this paper we we will mainly look at the ZX  sector.  So  we can choose the simplest model where
  \be
N_{\rm EI} =
  N_{\rm Special} 
 \la{simpexamp}\ee 
 
The quarks and leptons do not get mass splitting at tree level in any of these models.  If we start to look at mass splitting for the quarks and leptons, which happens at one loop level, we may need to reconsider this choice. 

\subsection{The Generator for the \Sei:}
\la{generatorsection}

The   counting operator $N_{\rm Special}$ in (\ref{nspecial1}) generates an invariance of the superpotential for the SSM.  Note the important facts that:
\ben
\item
It does not contain any term of the form  $J  \fr{\pa }{\pa J}$ that counts  the singlet J field, nor does it contain the combination $(N_H+
N_K)$. 
\item
It contains the H and K field counters with opposite signs.
\item
To be an invariant, the  other counters in  (\ref{nspecial1})  must also occur with the given signs, though we could change those parts somewhat by adding the invariants discussed above in (\ref{Nlepton}) and (\ref{Nquark}).
.
\item
This special invariant survives the generation of \sbgs.  It does so because of the absence of $J  \fr{\pa }{\pa J}$ and having the H and K field counters with opposite signs. 
\een
Recall that
\be
N_{\rm Special} =
  H  \fr{\pa }{\pa H }-
  K  \fr{\pa }{\pa K}
 +T  \fr{\pa }{\pa T}
- 
  B  \fr{\pa }{\pa B}-
 P  \fr{\pa }{\pa P}
+ 
  R  \fr{\pa }{\pa R}
\la{nspecial2}
 \ee

It is easy to see that this satisfies the constraint before we perform the shift to remove the VEVs:
\be
N_{\rm Special} \cP= \lt \{  H  \fr{\pa }{\pa H }-
  K  \fr{\pa }{\pa K}
 +T  \fr{\pa }{\pa T}
- 
  B  \fr{\pa }{\pa B}-
 P  \fr{\pa }{\pa P}
+ 
  R  \fr{\pa }{\pa R}
   \rt \}
   \eb
    \lt \{t Q K T +r L K R+b Q H B+p L H P
+g_{H} \fr{1}{\sqrt{2}} H K  J
-
g_{H} \fr{1}{\sqrt{2}} m^2 J 
\rt\}=0
\ee
It is also easy to see that this is also true for the theory where the shift has been performed, to implement spontaneous breaking of gauge symmetry:
 \be
N'_{\rm Special} \cP'= 0
\ee
which is \be
N'_{\rm Special} \cP'= \lt \{(H+ mh)  \fr{\pa }{\pa H }-
 (K+ mk)  \fr{\pa }{\pa K}
\ebp +T  \fr{\pa }{\pa T}
- 
  B  \fr{\pa }{\pa B}-
 P  \fr{\pa }{\pa P}
+ 
  R  \fr{\pa }{\pa R}
   \rt \}
   \eb
    \lt \{t Q  (K+ mk) T +r L (K+ mk) R+b Q (H+ mh) B\ebp +p  L (H+ mh) P
+g (H+ mh) (K+ mk)  J
-
g m^2 J 
\rt\}=0
\ee
What this means is that the constraint equation is  true for the theory with or without gauge symmetry breaking. As mentioned above, this is true because of the absence of the counting terms $n_1 (N_H+
N_K)  + n_2 N_J$ in $N_{\rm Special}$ in (\ref{nspecial2}). 

This yields the \EI \ generator:
\[
\cE_{\rm Special} = \Ct \f^{\dot \a}  \lt \{
(H +m h) \oy_{H \dot \a} -
(K +m k)  \oy_{K \dot \a} 
\rt. \]\be \lt.
 +T \oy_{T \dot \a} 
 -B \oy_{B \dot \a} -
 P  \oy_{P \dot \a} 
+ 
  R  \oy_{R \dot \a} 
\rt \}
\la{Especialgenerator}
\ee
 These generators and the form of the corresponding \EI s have been discussed in the papers \ci{E1,E2,E3,E4,E5}. This will be used below to construct the XM theory.
 
  The present $\cE_{\rm Special}$ is new, and it is the only known example where the \EI\ survives the \sbgs\ and generates tree level mass splitting for some mass-degenerate SUSY multiplets.  
As we shall see, the splitting of SUSY is a function of the VEV mass $m_{\rm VEV}$ that yields the \sbgs. 

 \section{Terms of the Action needed to create the XM}
 \la{newtermsforXM}
 
 Now we have established that we want to add the \EI\ that corresponds to the generator in equation (\ref{Especialgenerator}).  That has six parts, one for each of the  six terms in that generator.  The constraint equation ensures that these six parts will work together to make the whole object invariant under the BRS transformations, because the various parts that come from the superpotential will cancel with each other as they get generated by the pseudofields in the six parts.  This was illustrated in \ci{E4} for the simple case of the EP model, which is generated by a generator with only two parts.

 The action for the XM is the sum of the action for the SSM plus the action for the EI generated by (\ref{Especialgenerator}).
 
 We will return to the action for the SSM later.  Here we will look at the new EI, which is what is added to the SSM to make the XM.
 
 To prepare for the paper \ci{E7}, we really are interested only in two of the six parts of the \EI\ here.  These are the parts generated by $ g_{x}  \Ct \f^{\dot \a} (H^i+ m h^i) \oy_{H i\dot \a}$ and
 $- g_{x} 
\Ct \f^{\dot \a} (K^i+ m k^i) \oy_{K i\dot \a}  $.  Here they are in detail:

 \subsection{The  part   $ \cA_{\rm X \;Special, H }$ of the XM }

 \la{SSMHX}
  
\be
 g_{x}  \Ct \f^{\dot \a} (H^i+ m h^i) \oy_{H i\dot \a}  \ra
\ee
\be
 \cA_{\rm  X \;Special, H } = \eb
 +  g_{x}  \int d^4 x  \lt \{  \f^{\dot \a}\lt [ \lt ( 
b_{1}  (H^i+ m h^i)  \G_{H i} 
+ b_{2}\y_{H}^{ i \b }  Y_{Hi\b}
+b_{3}  F_{H}^{i} \Lam_{H i} 
\rt ) \oC_{ \dot \a}
\ebpp + 
b_{4} \y_{H }^{ i\a}
\lt ( \d^{j}_{ i} \pa_{\a \dot \a}  -\fr{1}{2}ig_{1} \d^{j}_{ i} V_{\a \dot \a}+ \fr{1}{2}i g_{2}  V^a_{\a \dot \a } \s^{aj}_{i} \rt ) 
 (\oH_j + m \oh_j) 
\ebpp
- i
b_{4}  (H^i+ m h^i) 
  \fr{1}{2} g_{1} \ov \lam_{\dot \a} (\oH_i+ m \oh_i)  
  \ebpp 
+ i b_{4}  (H^i+ m h^i) 
  \fr{1}{2} g_{2} \ov \lam^a_{\dot \a}
\s^{ai}_{j}
  (\oH_j + m \oh_j) 
 + 
b_{5} F_{H}^{i} \oy_{H i\dot \a}\rt ]
 \ebp  
+ X^{\dot \a \b} \lt [ \lt (b_{6}    (H^i+ m h^i)  Y_{H i \b} 
+ b_{7} \y^{i}_{H \b} \Lam_{H i}\rt )
\oC_{ \dot \a}+ 
b_{8} \y_{H \b}^{i}\oy_{H i  \dot \a} + 
\ebpp
 b_{9}   (H^i+ m h^i)
\lt ( \d^{j}_{ i} \pa_{\b \dot \a} 
+ \fr{1}{2} i g_{2}  V^a_{\b \dot \a } \s^{aj}_{i}  -\fr{1}{2}ig_{1} \d^{j}_{ i} V_{\b \dot \a}\rt )   
(\oH_{j}+m \oh_j)   \rt ]
\la{mixXandV}
 \ebp +  \c^{\dot \a} \lt [b_{10}   (H^i+ m h^i)   \Lam_{H i} \oC_{ \dot \a} +
b_{11}  (H^i+ m h^i)   \oy_{H i \dot \a}
 \rt ] \rt \}  
\la{mixchicandXH} 
\ee
As before in \ci{E4,E5} we still have:
\be
b_{1}=
b_{2}=
b_{3}=
b_{4}=b_{7}=b_{10}=1
;b_{5}=
b_{6}=
b_{8}=
b_{9}=
b_{11}= -1
\ee
\subsection{The  part   $ \cA_{\rm X \;Special, K }$ of the XM }

\la{SSMKX}

Note that the overall sign here comes from the fact that our counting operator in (\ref{nspecial2}) has opposite signs for $N_H$ and $N_K$:

 \be
- g_{x} 
\Ct \f^{\dot \a} (K^i+ m k^i) \oy_{K i\dot \a}  \ra
\ee
\be
  \cA_{\rm  X \;Special, K }
  \eb
  = - g_{x}  \int d^4 x  \lt \{  \f^{\dot \a}\lt [ \lt ( 
b_{1}  (K^i+ m k^i)  \G_{K i} 
+ b_{2}\y_{K}^{ i \b }  Y_{Ki\b}
+b_{3}  F_{K}^{i} \Lam_{K i} 
\rt ) \oC_{ \dot \a}
\rt.\rt.\eb\lt. \lt. + 
b_{4} \y_{K }^{ i\a}
\lt ( \d^{j}_{ i} \pa_{\a \dot \a}  +\fr{1}{2}ig_{1} \d^{j}_{ i} V_{\a \dot \a}
+ \fr{1}{2}ig_{2}  V^a_{\a \dot \a } \s^{aj}_{i} \rt ) 
 (\oK_j + m \ok_j) 
\rt. \ebp
i b_{4}  (K^i+ m k^i) 
  \fr{1}{2} g_{1} \ov \lam_{\dot \a} (\oK_i+ m \ok_i)   
\ebp+ib_{4}  (K^i+ m k^i) 
  \fr{1}{2} g_{2} \ov \lam^a_{\dot \a}
\s^{ai}_{j}
  (\oK_j + m \ok_j) + 
b_{5} F_{K}^{i} \oy_{K i\dot \a}\rt ]
 \eb\lt.  
+ X^{\dot \a \b} \lt [ \lt (b_{6}    (K^i+ m k^i)  Y_{K i \b} 
+ b_{7} \y^{i}_{K \b} \Lam_{K i}\rt )
\oC_{ \dot \a}+ 
b_{8} \y_{K \b}^{i}\oy_{K i  \dot \a} + 
\rt.\rt. \ee \be\lt. \lt. b_{9}   (K^i+ m k^i) 
\lt ( \d^{j}_{ i} \pa_{\b \dot \a} +\fr{1}{2}ig_{1} \d^{j}_{ i} V_{\b \dot \a}+ \fr{1}{2} i  g_{2} V^a_{\b \dot \a } \s^{aj}_{i}  \rt )   
  (\oK_j + m \ok_j) \rt ]
\la{KXmixing} \ebp +  \c^{\dot \a} \lt [b_{10}   (K^i+ m k^i)   \Lam_{K i} \oC_{ \dot \a} +
b_{11}  (K^i+ m k^i)   \oy_{K i \dot \a}
 \rt ] \rt \}  
\la{chipsiKmixing} 
\ee

As before in \ci{E4,E5} we still have:
\be
b_{1}=
b_{2}=
b_{3}=
b_{4}=b_{7}=b_{10}=1
;b_{5}=
b_{6}=
b_{8}=
b_{9}=
b_{11}= -1
\ee

The form (\ref{Especialgenerator}) implies that we also need more terms like the following for the various quarks and leptons:
\be
\Ct \f^{\dot \a} R^{p}   \oy_{R p \dot \a}  \ra
\ee
\be
 \cA_{\rm X, R } =  \int d^4 x  \lt \{  \f^{\dot \a}\lt [  \lt (
b_{1} R^{p}   \G_{R p} 
+ b_{2}\y^{p  \b }_{R }   Y_{R p \b} 
+b_{3}  F_{R}^{ p } \Lam_{R p}^{ } \rt)
  \oC_{ \dot \a}
  \la{axraction}\ebpp
 b_{4}  \y_{R }^{ p \a}
  \pa_{ \;\a \dot \a}
\ov R_{q}^{ } + b_{5} F_{R}^{p} \oy^{}_{R p\dot \a} 
\rt ]\rt. 
 \ee
 \be 
  + X^{\dot \a \b}  \lt [ \lt (b_{6}    R^{p}_{} Y_{R p  \b}^{} 
+ b_{7} \y^{p}_{R  \b} \Lam_{R p }^{}  \rt )
\oC_{ \dot \a}+ \ebp
b_{8} \y_{R \b }^{p}\oy^{}_{R p  \dot \a} + 
b_{9}    R^{p}_{} 
   \pa_{ \;\b \dot \a}  
\ov R_{p}^{ } \rt ]
 \ee
\be
 \lt. +  \c^{\dot \a} \lt [b_{10}  R^{p}_{}  \Lam_{R p}^{ } \oC_{ \dot \a} +
b_{11} R^{p}_{}  \oy^{}_{R p\dot \a}\rt ]\rt \}
\la{chiRovpsiR}
\ee
However a glance at this form shows that there is no mass term generated by this kind of term, because it does not contain the fields $H,K$.  So we can ignore these terms for this paper, since our concern here is with the ZX sector.

 \subsection{The Completion Terms for the Special Exotic Invariant, generalized from the EP Case}

The origin of these  can be found in \ci{E4,E5}.  They are also necessary for the preparation for \ci{E7}.

\be
\cA_{2,H}= g_x^2 \int d^4 x \lt\{ 
 \fr{1}{2}
\f^{\dot \a}
  \f_{\dot \a}    
  F^{i}_{H} \oH_i
  -\f_{\dot \a}
  X^{\dot \a \b}    
  \y^{i}_{H \b} \oH_{i}
\ebp
  + \fr{1}{2}X^{\dot \a \b}   X_{\dot \a \b}    H^{i}   \oH_{i}    + \f^{\dot \a }   \c_{\dot \a}    H^{i}   \oH_{i}  
\rt \} +*
\ee

\be
 \cA'_{2,H}=   \int d^4 x \lt \{
s_{1}  \f^{\dot \a}  \pa_{\a \dot \a} \ov \f^{\a}
  H^{i}   \oH_{i}    +s_{2}
  \f^{\dot \a}    \ov X_{\a \dot \a}
 \y_{H}^{i \a}     \oH_{i}     
  \ebp
  +s_{3}
  \ov \f^{ \a}     X_{\dot \a \a} 
 H^{i}     \oy_{H i}^{\dot \a}  
    +s_{4}
 \f_{\dot \a}  \ov \f_{\a} 
   \oy_{H i}^{\dot \a} 
 \y_{H }^{i\a} 
     +  s_{5}
      X_{\dot \a \a}   \oX^{\dot \a \a}   H^{i}   \oH_{i} 
 \rt \}
\ee
When the fields $H,K$ are shifted by the VEV, these terms will give rise to mass terms for the various CDSS fields.  We will return to them in \ci{E7}.

Similar pieces for the doublets $\cA_{2,K},\cA'_{2,K}$ are also present of course, and for the Right Singlets, such as
$\cA_{2,J},\cA'_{2,J}$.  

There are also terms like these for the quarks and leptons, but again those give rise to no mass terms.  These are $\cA_{2,L},\cA'_{2,L}$ and $\cA_{2,Q},\cA'_{2,Q}$ for the two Matter Left Doublets and  $\cA_{2,P},\cA'_{2,P}$ for t  and the four Matter Right Singlets:
\be
\cA_{2, P}= g_x^2 \int d^4 x \lt\{ 
 \fr{1}{2}
\f^{\dot \a}
  \f_{\dot \a}    
  F_{P} \oP
  -\f_{\dot \a}
  X^{\dot \a \b}    
  \y_{P \b} \oP
\ebp
  + \fr{1}{2}X^{\dot \a \b}   X_{\dot \a \b}    P   \oP  + \f^{\dot \a }   \c_{\dot \a}    P   \oP  
\rt \} 
\ee

\be
 \cA'_{2.P}=   \int d^4 x \lt \{
s_{1}  \f^{\dot \a}  \pa_{\a \dot \a} \ov \f^{\a}
 P \oP +s_{2}
  \f^{\dot \a}    \ov X_{\a \dot \a}
  \y_{P}^{\a}  \oP 
  \ebp
  +s_{3}
  \ov \f^{ \a}     X_{\dot \a \a} P
  \oy_{P}^{\dot \a}  
   +s_{4}
 \f_{\dot \a}  \ov \f_{\a} \oy_P^{\dot \a}  \y_P^{\a}
    +  s_{5}
      X_{\dot \a \a}   \oX^{\dot \a \a}  P \oP 
 \rt \}
\ee
We can ignore these for present purposes.

 \subsection{The \CDSS}
 \la{CDSSsection}
 
Here are the actions for these fields.  Note that there are two different kinds of kinetic actions for these fields.  That will be important in \ci{E7}.

 \be
\cA_{\rm CDSS,\;Kinetic}    
= a_1 \int d^4 x   
\lt \{  {\f}_{\dot \b} \pa^{\a \dot \b} \Box {\ov \f}_{\a} 
+ {X}_{\dot \b\a } \pa^{\a \dot \d} \pa^{\g \dot \b}   {\ov S}_{\g \dot \d} 
+  {\c}_{\dot \b} \pa^{\a \dot \b}  {\ov \c}_{\a}  
\rt \} \ee

\be \cA_{\rm CDSS, \;Chiral}  =    \int d^4 x   
\lt \{2  \f^{\dot \b} ( a_2\Box  +m_x^2) \c_{\dot \b}  
 +2    X^{ \dot \b \a}  ( a_2\Box  +m_x^2 )  X_{\dot \b \a }
\rt \} + *
\ee

\section{A Quick Look at the SSM action}
\la{SSMaction}
These terms will be needed for the next paper \ci{E7}, which will explore the mass terms that split the SUSY multiplets in the ZX  sectors at tree level. This part of the action for the SSM is well known. 

\subsection{Gauge Actions}

In this paper we will usually ignore the quarks and the SU(3) gauge fields.  

\be {\cal A}_{\rm SUSY\;Gauge\; SU(2) } = 
\int d^{4}x \left \{ -\frac{1}{4} 
 \lt ( \pa_{\m} V^a_{\n} 
-\pa_{\n} V^a_{\m} 
+  g_{2} \ve^{abc} V^b_{\m} V^c_{\n} \rt )
 \ebp 
 \lt ( \pa^{\m} V^{a \n} 
-\pa^{\n} V^{a \m} 
+  g_{2}  \ve^{ade} V^{d\m} V^{e \n} \rt )
\ebp
 - \frac{1}{2}  \lambda^{a \a}
 \lt ( \pa_{\a \dot \b}  {\ov  \lambda}^{a \dot \b} 
+
 g_{2}  \ve^{abc} V_{\a \dot \b}^{b} {\ov  \lambda}^{b \dot \b}\rt )
 + \frac{1}{2}
D^{a}D^{a} \rt \}
 \ee
 
\be 
{\cal A}_{\rm  SUSY\;Gauge\;  U(1) } = 
\int d^{4}x \left \{ -\frac{1}{4} 
\lt (\pa_{\m} U_{\n} 
-\pa_{\n} U_{\m} \rt ) \lt ( \pa^{\m} U^{ \n} 
-\pa^{\n} U^{ \m} \rt )
\ebp - \frac{1}{2}  \lambda^{ \a}
 \pa_{\a \dot \b}  {\ov  \lambda}^{ \dot \b} 
 + \frac{1}{2}
D D  \rt \}
\ee

\subsection{\rm Faddeev--Popov Ghosts, Gauge Fixing and Ghost Action}
 
\be 
{\cal A}_{\rm Ghost\;and\; Gauge\; Fixing} ={\cal A}_{\rm GGF} =\eb
  \d  \int d^{4}x \;  \h^{a} \left \{ \frac{1}{2} Z^{a} + \fr{1}{2 \a_{\rm g}}\partial^{\mu}V_{\mu}^{a}
  + 2 \a_{\rm g} G^a
\right \}
\ee

\subsection{${\cal A}_{\rm ZJSYM, U(1)}$ Action}

The  Z vector boson and its superpartners are in the ZX sector and we will need to include their transformtions in \ci{E7}:

\be
{\cal A}_{\rm ZJSYM, U(1)} = \int d^4 x \;
\ee
\be
\left \{ 
\S^{ \a \dot \b } 
\lt [ \pa_{\a \dot \b} \w
  + 
  C_{\a} 
\ov{\lambda}_{ \dot{\b} }   +
  \lambda^{}_{\a } 
{\ov C}_{\dot{\b} }   +  \x^{\g \dot \d}\pa_{\g \dot \d}V_{\a \dot \b} \rt ]
\ebp
 +
 L^{ \a} 
\lt[  \frac{1}{2}\lt (
 \pa_{\a \dot \b} V_{\b}^{\;\; \dot \b}
+ \pa_{\b \dot \b} V_{\a}^{\;\; \dot \b}
 \rt )
   C^{\b}   - i D^{}
C_{\a}  +  \x^{\g \dot \d}\pa_{\g \dot \d}
\lambda^{}_{\a} 
 \rt ]
	\la{variationoflambda1}
		\ebp 
+ \ov{L}^{ \dot \a}  
\lt [\frac{1}{2}  \lt (
 \pa_{\g \dot \a} V^{\g}_{\;\; \dot \b}
+  \pa_{\g \dot \b} V^{\g}_{\;\; \dot \a}
 \rt )
\ov{C}^{\dot \b}     +  i D^{}  {\ov C}_{ \dot \a}  +
 \x^{\g \dot \d}\pa_{\g \dot \d}\ov{\lambda}^{}_{\dot \a}  
\rt ]
\ebp
+
\D^{}
\lt [\fr{-i}{2}  C^{\a} 
  \pa_{\a \dot \b}  {\ov  \lambda}^{ \dot \b} 
  +\fr{i}{2}\ov{C}^{\dot \b}
  \pa_{\a \dot \b}  { \lambda}^{ \a} 
+
 \x^{\g \dot \d}\pa_{\g \dot \d} D^{}
\rt ]
\ebp
+
W^{}
\lt [
  C^{\a} \ov{C}^{\dot \b}  V_{\a
\dot \b} +
 \x^{\g \dot \d}\pa_{\g \dot \d} \omega\rt ]
\ebp
+
H^{} 
\lt ( Z+  \x^{\g \dot \d}\pa_{\g \dot \d}  \h 
\rt )
\ebp+
\z
\lt ( C^{\a}  
\ov{C}^{\dot \b}
\pa_{\a \dot \b} \h +  \x^{\g \dot \d}\pa_{\g \dot \d}Z 
\rt )
\right \}
\ee
 
\subsection{${\cal A}_{\rm ZJSYM, SU(2)}$ Action}

\be
{\cal A}_{\rm ZJSYM,SU(2)} = \int d^4 x \;
\ee
\be
\left \{ 
\S^{a \a \dot \b } 
\lt ( 
\lt ( \pa_{\a \dot \b} \w^{a}
+
g_{2} \ve^{abc} V_{\a \dot \b}^{b} \w^{c}
\rt ) + 
  C_{\a} 
\ov{\lambda}^{a}_{ \dot{\b} }   +
  \lambda^{a }_{\a } 
{\ov C}_{\dot{\b} }   + \x^{\nu}
\partial_{\nu} V_{\a \dot \b}^{a} \rt )
\ebp
 +
 L^{a \a} 
  \lt (
   \frac{1}{2}
 \lt ( \pa_{\m} V^a_{\n} 
-\pa_{\n} V^a_{\m} 
+   g_{2}  ve^{abc}  V^b_{\m} V^c_{\n} \rt )\s^{\mu \nu}_{\a \b} C^{\b} 
\ebpp  + i g_{2}  \ve^{abc}  \lambda^{b}_{\a} \omega^{c} - i D^{a}
C_{\a}  + \x^{\nu} \partial_{\nu}
\lambda^{a}_{\a} 
 \rt )
	\la{variationoflambda2}
		\ebp 
+ \ov{L}^{a \dot \a}  
\lt (  \frac{1}{2}
 \lt ( \pa_{\m} V^a_{\n} 
-\pa_{\n} V^a_{\m} 
+   g_{2}  \ve^{abc}  V^b_{\m} V^c_{\n} \rt ) \ov{\s}^{\mu \nu}_{\dot{\a} 
\dot{\b} } 
\ov{C}^{\dot \b} 		\ebpp 
 -i   g_{2} \ve^{abc}  \ov{\lambda}^{b}_{\dot \a} \omega^{c}  +  i D^{a}  {\ov C}_{ \dot \a}  +
\x^{\nu} \partial_{\nu}\ov{
\lambda}^{a}_{\dot \a}  
\rt )
\ebp
+
W^{a}
\lt [
 - \frac{1}{2} g_{2} \ve^{abc} 
\omega^{b} \omega^{c} + C^{\a} \ov{C}^{\dot \b}  V_{\a
\dot \b}^a +
\x^{\nu} \partial_{\nu} \omega^{a}\rt ]
\ebp
+
H^{a} 
\lt ( Z^a + \x^{\nu} \partial_{\nu}  \h^a 
\rt )
\ebp+
\z^{a}
\lt ( C^{\a} \s^{\mu}_{\a \dot{\b} }
\ov{C}^{\dot \b}
\pa_{\m} \h^a + \x^{\m} \pa_{\m} Z^a 
\rt )
\ebp
+
\D^{a}
\lt [\fr{-i}{2}  C^{\a} 
\lt ( \pa_{\a \dot \b}  {\ov  \lambda}^{a \dot \b} 
+
  g_{2} f^{abc}   V_{\a \dot \b}^{b} {\ov  \lambda}^{c \dot \b}
\rt ) 
\ebpp
+\fr{i}{2}\ov{C}^{\dot \b}
\lt ( \pa_{\a \dot \b}  { \lambda}^{c \a} 
+
 g_{2} \ve^{abc}  V_{\a \dot \b}^{b} { \lambda}^{c \a}\rt )
\ebpp
 + i  g_{2} \ve^{abc}  D^{b} \omega^{c}  +
\x^{\nu} \partial_{\nu} D^{a}
\rt ]
\right \}
\ee

 \subsection{\rm The subactions  ${\cal A}_{\rm SSM,H}$  and ${\cal A}_{\rm SSM,K}$ }

\la{SSMHK}

There are also mass terms for the ZX  sector generated in these two parts of the action after gauge symmetry breaking.  These will mix with the mass terms in section \ref{newtermsforXM}, and that is the problem that will be looked at in \ci{E7}. 

\be {\cal A}_{\rm SSM,H} = 
\int d^{4}x \left \{
  F_H^{ i}    {\ov F}_{H i } 
  \ebp
  +
\y^{i  \a  }_{H} \lt ( \pa_{ \a \dot \a  }  {\ov \y}_{H i }^{  \dot \a}
- i    \fr{1}{2} g_{1}  V_{ \a \dot \a }   {\ov \y}_{H j }^{  \dot \a} +i g_{2}  \fr{1}{2}\s^{a j}_{\;\;\;\;i} V^a_{ \a \dot \a }   {\ov \y}_{H j }^{  \dot \a}\rt )
\ebp   
+
\lt ( \pa_{ \a \dot \a  }  \d^j_i+ i    \fr{1}{2}  \d^j_i g_{1}  V_{ \a \dot \a }  
- i g_{2}  \fr{1}{2}\s^{a j}_{\;\;\;\;i} V^a_{ \a \dot \a }  \rt )  (H^{i}+ m h^i)
\ebp\lt ( \pa^{ \a \dot \a  }  \d^k_j - i    \fr{1}{2} g_{1} \d^k_j V_{ \a \dot \a }
+i g_{2}  \fr{1}{2}\s^{a k}_{\;\;\;\;j} V^a_{ \a \dot \a }   \rt )  (\oH_{k} + m \oh_k)
\ebp
-    g_{1}  \fr{1}{2} 
  \left (     D  (H^j +
m h^j )  (\oH_j +
m \ov h_j ) 
\rt.\ebp
+    g_{2}  \fr{1}{2}  \s^{a i}_{\;\;\;\;j} 
  \left (     D^a  (H^j +
m h^j )  (\oH_j +
m \ov h_j ) 
\rt.
\ebp   
+ i g_{1}
\oy_{H i }^{\dot\a  }  \ov \lambda_{\dot \a  } (H^i +
m h^i ) 
-  i g_{2}  \fr{1}{2}  \s^{a i}_{\;\;\;\;j}
{\ov \y}_{H i}^{  \dot \b}
{\ov \lambda}^{ a}_{ \dot \b}    (H^j +
m h^j )   
\ebp
- i g_{1}
\y_H^{ i \a  }  \lambda_{\a  } (\oH_i +
m \ov h_i )
+  i g_{2}  \fr{1}{2}  \s^{a i}_{\;\;\;\;j}
{ \y}_{H }^{j  \b}
{\lambda}^{ a}_{   \b}    (\oH_i +
m \oh_i )   
\rt\}
\ee

\be {\cal A}_{\rm SSM,K} =  
\int d^{4}x \left \{
  F_K^{ i}    {\ov F}_{K i } 
  \ebp
  +
\y^{i  \a  }_{K} \lt ( \pa_{ \a \dot \a  }  {\ov \y}_{K i }^{  \dot \a}
- i g_{2}  \fr{1}{2}\s^{a j}_{\;\;\;\;i} V^a_{ \a \dot \a }   {\ov \y}_{K j }^{  \dot \a}- i    g_{1} \fr{1}{2}  V_{ \a \dot \a }   {\ov \y}_{K j }^{  \dot \a}\rt )
\ebp   
+
\lt ( \pa_{ \a \dot \a  }  \d^j_i
+ i g_{2}  \fr{1}{2}\s^{a j}_{\;\;\;\;i} V^a_{ \a \dot \a }  + i    \fr{1}{2}  \d^j_i g_{1}  V_{ \a \dot \a }  \rt )  (K^{i}+ m k^i)
\ebp\lt ( \pa^{ \a \dot \a  }  \d^k_j
-i g_{2}  \fr{1}{2}\s^{a k}_{\;\;\;\;j} V^a_{ \a \dot \a }  - i    \fr{1}{2} g_{1} \d^k_j V_{ \a \dot \a }  \rt )  (\oK_{k} + m \ok_k)
\ebp
+    g_{1}  \fr{1}{2} 
  \left (     D  (K^j +
m k^j )  (\oK_j +
m \ov k_j ) 
\rt.+    g_{2}  \fr{1}{2}  \s^{a i}_{\;\;\;\;j} 
  \left (     D^a  (K^i +
m k^i )  (\oK_j +
m \ov h_j ) 
\rt. 
\ebp
- i g_{1}\fr{1}{2} 
\oy_{K i }^{\dot\a  }  \ov \lambda_{\dot \a  } (K^i +
m k^i ) -  i g_{2}  \fr{1}{2}  \s^{a i}_{\;\;\;\;j}
{\ov \y}_{K i}^{  \dot \b}
{\ov \lambda}^{ a}_{ \dot \b}    (K^j +
m k^j )   
\ebp 
 + i g_{1}\fr{1}{2} 
\y_K^{ i \a  }  \lambda_{\a  } (\oK_i +
m \ov h_i ) 
 +  i g_{2}  \fr{1}{2}  \s^{a i}_{\;\;\;\;j}
{ \y}_{K }^{j  \b}
{\lambda}^{ a}_{   \b}    (\oK_i +
m \ok_i )   
\rt\}
\la{psiKlamoK}\ee

 \subsection{\rm The transformations for the H component fields in the  SSM }
\la{SSMZJH}
 
 This summarizes the transformation for the fields in the H multiplet.  The action for the K multiplet is similar but it reflects the different quantum numbers of course.

\be
{\cal A}_{\rm SSM, ZJWZ, H} = \int d^4 x \;
\ee
\be
\left \{ 
\G_{ Hi}    \lt [ \y^{i}_{ H \b} {  C}^{  \b} 
- \fr{1}{2} i g_{1} \w H^i +\fr{1}{2}   i g_{2} \s^{a i}_{\;\;\;j} \w^a H^j
+ \x^{\nu} \partial_{\nu}   H^i \rt ]
\ebp
+
 Y_{i}^{ \a} \lt [
\lt ( \pa_{\a \dot \b}  H^i +\fr{1}{2}  i g_{1}   V_{\a \dot \b}H^i 
-   \fr{1}{2} i g_{2} \s^{a i}_{\;\;\;j}
 V^a_{\;\;\a \dot \b}  H^j \rt ) {\ov C}^{\dot \b}   
\ebpp
+ F_{H}^{i} {C}_{ \a} - \fr{1}{2} i g_{1} \w \y^{j}_{H \a  }  +   \fr{1}{2}i g_{2} \s^{a i}_{\;\;\;j} \w^a \y^{j}_{H \a  } 
 + \x^{\nu} \partial_{\nu}  \y^{i}_{H \a  }\rt ]
\ebp
+\Lam_i  \lt [{\ov C}^{\dot \b}
\lt ( \pa_{\a \dot \b}  \y_{H }^{i \a}  +\fr{1}{2}  i g_{1} ( V_{\a \dot \b}   \y_{H }^{i \a} + {\ov \lambda}_{ \dot \b}  H^i  )
\rt.\ebpp\lt.
-   \fr{1}{2} i g_{2} \s^{a i}_{\;\;\;j} ( V^a_{\;\;\a \dot \b}     \y_{H }^{j \a}  + {\ov \lambda}^{a}_{ \dot \b}  H^i  ) \rt )
\ebpp- \fr{1}{2} i g_{1} \w F_H^i +\fr{1}{2}   i g_{2} \s^{a i}_{\;\;\;j} \w^a F_H^j
 + \x^{\nu} \partial_{\nu}  F_H^i \rt ]
\rt \}
\ee

\be
{\cal A}_{\rm ZJStructure} =  -{P}^{\a \dot \b}    C_{\a} {\ov C}_{ \dot \b}
\ee

 \subsection{\rm $ {\cal A}_{\rm SSM,J}$ and   ${\cal A}_{\rm SSM,L}$ and ${\cal A}_{\rm SSM,R}$ etc.}
No masses originate in these, so we can ignore them for now:
 
\be {\cal A}_{\rm SSM,J} = 
\int d^{4}x \left \{
  F_J     {\ov F}_{J } 
  +
\y^{ \a  }_{J}   \pa_{ \a \dot \a  }  {\ov \y}_{J }^{  \dot \a} 
+
  \pa_{ \a \dot \a  }  J
  \pa^{ \a \dot \a  }\oJ 
\rt\}
\ee
 
\be {\cal A}_{\rm SSM,L} = 
\int d^{4}x \left \{
  F_L^{ i p}    {\ov F}_{L i p} 
  \ebp
  +
\y^{i p \a  }_{L} \lt ( \pa_{ \a \dot \a  }  {\ov \y}_{L i p}^{  \dot \a}
- i g_{2}  \fr{1}{2}\s^{a j}_{\;\;\;\;i} V^a_{ \a \dot \a }   {\ov \y}_{L j p}^{  \dot \a}+ i    \fr{1}{2} g_{1}  V_{ \a \dot \a }   {\ov \y}_{L j p}^{  \dot \a}\rt )
\ebp   
+
\lt ( \pa_{ \a \dot \a  }  \d^j_i
- i g_{2} \fr{1}{2}\s^{a j} _{\;\;\;\;i} V^a_{ \a \dot \a }  + i    \fr{1}{2}  \d^j_i g_{1}  V_{ \a \dot \a }  \rt )  L^{ip}\ebp\lt ( \pa^{ \a \dot \a  }  \d^k_j
-i g  \fr{1}{2}\s^{a k}_{\;\;\;\;j} V^a_{ \a \dot \a }  + i    \fr{1}{2} g_{1} \d^k_j V_{ \a \dot \a }  \rt )  \oL_{k p}  
\ebp
+    g_{2}  \fr{1}{2}  \s^{a i}_{\;\;\;\;j} 
     D^a  L^{jp}  \oL_{jp} 
-    g_{1}  \fr{1}{2} 
  D  L^{jp}   \oL_{jp}
\ebp   
 +  i g_{2}  \fr{1}{2}  \s^{a i}_{\;\;\;\;j}
{\ov \y}_{L i p}^{  \dot \b}
{\ov \lambda}^{ a}_{ \dot \b}    L^{jp} 
 -  i g_{2}  \fr{1}{2}  \s^{a i}_{\;\;\;\;j}
{ \y}_{L }^{j p \b}
{\lambda}^{ a}_{   \b}    \oL_{ip}  
\ebp 
- i g_{1}
\y_L^{ i \a  }  \lambda_{\a  } \oL_{ip}
+ i g_{1}
\oy_{L i }^{\dot\a  }  \ov \lambda_{\dot \a  } L^{ip}
\rt\}
\ee

\be {\cal A}_{\rm SSM,R} = 
\int d^{4}x \left \{
  F^q_R     {\ov F}_{R q} 
  +
\y^{ q \a  }_{R}   \pa_{ \a \dot \a  }  {\ov \y}_{R q}^{  \dot \a} 
+
  \pa_{ \a \dot \a  }  R^q
  \pa^{ \a \dot \a  }\oR_q 
\rt\}
\ee

\subsection{ $\cA_{\rm Superpotential, LKR}$}

At tree level all the masses of the quarks and leptons originate here and  $\cA_{\rm Superpotential, QHB}$ etc. The origin of the masses for the Z sector and the Higgs sector are a little more complicated, and we will discuss them in \ci{E7}.

\be
r_{pq} \Ct L^{ip} (K_{i} + m  k_i) R^{q} \ra 
\ee \be
\cA_{\rm Superpotential, LKR}=
\eb \int d^4 x \lt \{ 
r_{pq} L^{ip} (K_{i} + m  k_i) 
 F_{R}^{q}\ebp+r_{pq}  F_{L}^{ip}(K_{i} + m  k_i) R^{q} 
 +  r_{pq}L^{ip} F_{K i}  R^{q} 
 \ebp
-
r_{pq}L^{ip} \y_{ K i}^{\a}  \y_{R \a}^{q } 
-
r_{pq}\y_{L}^{ip\a} (K_{i} + m  k_i)\y_{R \a}^{q } 
\ebp
-
r_{pq}\y_{L}^{ip\a} \y_{ K i \a}   R^{q } 
\rt \}
\ee
 
The \CC\ of ${\cA}_{\rm Superpotential, LKR}$ is 

\be
{\ov\cA}_{\rm Superpotential, LKR}=
\eb \int d^4 x \lt \{ 
\ov r^{pq}\oL_{ip} (\oK^{i} + m  \ok^i) 
 \oF_{R q}\ebp+\ov r^{pq}  \oF_{L ip}(\oK^{i} + m  \ok^i) \oR_{q} 
 + \ov r^{pq} \oL_{ip} \oF_{K}^{ i}  \oR_{q} 
 \ebp
-
\ov r^{pq} \oL_{ip} \oy_{ K i}^{\dot \a}  \oy_{R q\dot \a } 
-
\ov r^{pq} \oy_{L ip\dot \a}(\oK^{i} + m  \ok^i) \oy_{R a \dot \a} 
\ebp
-
\ov r^{pq} \oy_{L ip}^{\dot \a} \oy^{i}_{ K  \dot \a}   \oR_{q } 
\rt \}
\la{ovpsipsiR})\ee

\subsection{The \ME}
\la{mastereqsection}

The \ME\ has the following form: 
\be
\cM_{\rm }=\cM_{\rm H}+\cM_{\rm K}+\cM_{\rm J} 
+\cM_{\rm L} +\cM_{\rm P} +\cM_{\rm R} \eb +\cM_{\rm Q} +\cM_{\rm T} +\cM_{\rm B} 
+\cM_{\rm X}+\cM_{\rm Structure}=0
\ee
Here are three examples:
\be
\cM_{\rm H}=
\int d^4 x \lt \{
\fr{\d \cA}{\d H^i} \fr{\d \cA}{\d \G_{Hi}}
+
\fr{\d \cA}{\d \y^i_{H \a}  } \fr{\d \cA}{\d Y_{Hi}^{\a}}
+
\fr{\d \cA}{\d F_H^i} \fr{\d \cA}{\d \Lam_{Hi}}
+
\fr{\d \cA}{\d \oH_i} \fr{\d \cA}{\d \ov\G_H^i}
\ebp+
\fr{\d \cA}{\d \oy_{H i  \dot \a}} \fr{\d \cA}{\d \oY_H^{i\dot \a}}
+
\fr{\d \cA}{\d \oF_{Hi}} \fr{\d \cA}{\d \ov\Lam_H^i}
\rt \}  
\la{wzh}
\ee

\be
\cM_{\rm X} =\int d^4 x \lt \{
\fr{\d \cA}{\d \f_{\dot\a}} \fr{\d \cA}{\d  G^{\dot\a}}
+
\fr{\d \cA}{\d X_{\a\dot\a}} \fr{\d \cA}{\d \S^{\a\dot\a}}
+
\fr{\d \cA}{\d\c_{\dot\a}} \fr{\d \cA}{\d L^{\dot\a}}
 +\fr{\d \cA}{\d \ov \f_{\a}} \fr{\d \cA}{\d  \ov G^{ \a}}
\ebp+
\fr{\d \cA}{\d \ov X_{\dot\a\a}} \fr{\d \cA}{\d \ov\S^{\dot\a\a}}
+
\fr{\d \cA}{\d\ov\c_{\a}} \fr{\d \cA}{\d \ov L^{\a}}
\rt \}  
\la{mastercdssdownfield} 
\ee
\be
 \cM_{\rm Structure}=
\fr{\pa \cA}{\pa P_{\a \dot \b}} 
\fr{\pa \cA}{\pa \x^{\a \dot \b}}    
\la{masterstructure}
\ee
The importance of the above is discussed at length in the papers in the E  series: \ci{E1,E2,E3,E4,E5,E6,E7,E??}

 \section{Conclusion}
 
 This paper singles out the XM model for further examination.  This model has the merit that the introduction of an exotic invariant to this model survives the spontaneous breaking of gauge symmetry, and that  spontaneous breaking of gauge symmetry also gives rise to SUSY splitting mass mixing in the ZX  sector from the new parts of the action labelled $\cA_{\rm  X \;Special, H }$ and  $\cA_{\rm  X \;Special, K }$.  All these terms originate in section \ref{newtermsforXM},  which includes all the new terms for the XM that are not in the original SSM.
 We also include the usual terms for the SSM in section 
 \ref{SSMaction}.  These would give rise to degenerate mass multiplets if the terms in section \ref{newtermsforXM} were not present.

 The next logical step to be taken here is to work out the spectrum of the particles in the ZX sector after gauge symmetry breaking, at tree level.  The particles in the ZX sector are evident from the action in section \ref{newtermsforXM} above.  These are the Z vector boson and its superpartners consisting of two fermions and a scalar, and the CDSS X multiplet which consists of a complex vector boson and two spinors.  These get mixed by gauge symmetry breaking through the terms in section \ref{newtermsforXM}.
 It seems remarkable that the muliplets that get SUSY mass splitting at tree level are the neutral flavour--diagonal multiplets in the ZX  sector, so that it appears possible that this can explain the origin of the observed suppression of flavour changing neutral currents. The crucial question is whether the mass splitting that we find in \ci{E7} makes sense physically, particularly since the CDSS multiplet looks very tricky to deal with. It appears that the existence of the two independent kinetic terms in section \ref{CDSSsection} may lend some useful flexibility in that regard.

\begin{center}
 { Acknowledgments}
\end{center}
\vspace{.1cm}

  I thank  Carlo Becchi,  Margaret Blair, Friedemann Brandt, Philip Candelas, David Cornwell,   James Dodd, Mike Duff, Sergio Ferrara, Richard Golding, Dylan Harries, Marc Henneaux, Chris T.  Hill,  D.R.T. Jones, Olivier Piguet, Antoine van Proeyen,  Pierre Ramond,   Peter Scharbach,      Mahdi Shamsei, Kelly Stelle, Sean Stotyn, Xerxes Tata, J.C. Taylor,  Peter West and Ed Witten for stimulating correspondence and conversations.   I also express appreciation for help in the past from William Deans, Lochlainn O'Raifeartaigh, Graham Ross, Raymond Stora, Steven Weinberg, Julius Wess and Bruno Zumino. They are not replaceable and they are missed.  I   thank  Doug Baxter,  Margaret Blair,  Murray Campbell, David Cornwell, James Dodd, Davide Rovere,   Pierre Ramond, Peter Scharbach and Mahdi Shamsei for recent, and helpful, encouragement to carry on with this work.


 \tiny 
\articlenumber\\
\today
\hourandminute

\end{document}

%% file: Johnmacros29.tex
\newcommand{\Sei}{{Special Exotic Invariant for the SSM}}

\newcommand{\Ct}{{\rm (C \x^2 C)  }}

  \newcommand{\cdss}{chiral dotted spinor superfield}

\newcommand{\XM}{{Exotic Model}}
\newcommand{\vev}{vacuum expectation value}

\newcommand{\sbgs}{spontaneous breaking of gauge symmetry}

\newcommand{\ME}{Master Equation}

\newcommand{\cM}{{\cal M}}

\newcommand{\CDSS}{{Chiral Dotted Spinor Superfield}}

\newcommand{\cE}{{\cal E}}

\newcommand{\cA}{{\cal A}}

\newcommand{\CC}{Complex Conjugate}

\newcommand{\Exists}{\bm{\exists}\kern-0.6em\bm{\exists}}

\newcommand{\EI}{Exotic Invariant}

\newcommand{\SSM}{Supersymmetric Standard Model}

\newcommand{\sg}{$\rm SU(3) \times SU(2) \times U(1)$}

\newcommand{\bsg}{$\rm SU(3)  \times U(1)$}

\newcommand{\bt}{\begin{tabular}{c}}
\newcommand{\et}{\end{tabular}}

\newcommand{\eb}{\ee\be } 
\newcommand{\ebp}{\rt.\ee\be\lt.} 
\newcommand{\ebpp}{\rt.\rt.\ee\be\lt.\lt.} 
 
\newcommand{\bmat}{\lt ( \begin{array} }
\newcommand{\emat}{  \end{array} \rt )}

\newcommand{\oH}{{\ov H}}
\newcommand{\oP}{{\ov P}}

\newcommand{\oR}{{\ov R}}

\newcommand{\cP}{{\cal P}}

\newcommand{\oJ}{{\ov J}}

\newcommand{\oX}{{\ov X}}

\newcommand{\oK}{{\ov K}}

\newcommand{\oY}{{\ov Y}}

\newcommand{\oy}{{\ov \y}}

\newcommand{\oh}{{\ov h}}

\newcommand{\ok}{{\ov k}}
\newcommand{\oC}{{\ov C}}
\newcommand{\oL}{{\ov L}}

\newcommand{\oF}{{\ov F}}

\renewcommand{\a}{\alpha}	
\renewcommand{\b}{\beta}
\newcommand{\g}{\gamma}
\renewcommand{\d}{\delta}
\newcommand{\e}{\epsilon}
\newcommand{\ve}{\varepsilon}
\newcommand{\z}{\zeta}
\newcommand{\h}{\eta}

\newcommand{\lam}{\lambda}
\newcommand{\m}{\mu}
\newcommand{\n}{\nu}	
\newcommand{\x}{\xi}

\newcommand{\s}{\sigma}

\newcommand{\f}{\phi}
\renewcommand{\c}{\chi}
\newcommand{\y}{\psi}
\newcommand{\w}{\omega}
\newcommand{\G}{\Gamma}
\newcommand{\D}{\Delta}
	
\newcommand{\Lam}{\Lambda}

\renewcommand{\S}{\Sigma}

\newcommand{\la}{\label}
\newcommand{\ci}{\cite}

\newcommand{\ds}{\documentstyle}	
\newcommand{\fr}{\frac}

\newcommand{\pa}{\partial}
\newcommand{\ov}{\overline}
\newcommand{\br}{\begin{rant}}
\newcommand{\er}{\end{rant}}
\newcommand{\beC}{\begin{Conjecture}}
\newcommand{\eeC}{\end{Conjecture}}
\newcommand{\be}{\begin{equation}}
\newcommand{\ee}{\end{equation}}
\newcommand{\ba}{\begin{array}} 
\newcommand{\ea}{\end{array}}
\newcommand{\bea}{\begin{eqnarray}}
\newcommand{\eea}{\end{eqnarray}}
\newcommand{\ra}{\rightarrow}

\newcommand{\lt}{\left}
\newcommand{\rt}{\right}

\newcommand{\ben}{\begin{enumerate}}
\newcommand{\een}{\end{enumerate}}
\newcommand{\bitem}{\begin{itemize}}
\newcommand{\eitem}{\end{itemize}}